# New high-pressure form of boron is significantly ionic.

## Reply to the rejected comment by Dubrovinskaia et al.
(*http://arxiv.org/abs/0907.1900*)

Artem R. Oganov[1,2,10], Jiuhua Chen[3,4], Carlo Gatti[5], Yanzhang Ma[6], Yanming Ma[1,7], Colin W. Glass[1], Zhenxian Liu[8], Tony Yu[3], Oleksandr O. Kurakevych[9], Vladimir L. Solozhenko[9]

[1] *ETH Zurich, Wolfgang-Pauli-Str. 10, CH-8093 Zurich, Switzerland.*
[2] *Moscow State University, 119992 Moscow, Russia.*
[3] *Florida International University, Miami, FL 33199, USA*
[4] *Stony Brook University, Stony Brook, NY 11794-2100, USA*
[5] *CNR-ISTM Istituto di Scienze e Tecnologie Molecolari, via Golgi 19, 20133 Milano, Italy.*
[6] *Texas Tech University, 7th St. & Boston Ave., Lubbock, Texas 79409, USA.*
[7] *Jilin University, Changchun 130012, P. R. China.*
[8] *Geophysical Laboratory, Carnegie Institution of Washington, Washington, DC 20015, USA*
[9] *LPMTM-CNRS, Université Paris Nord, Villetaneuse, F-93430, France*
[10] *Present address: Stony Brook University, Stony Brook, NY 11794-2100, USA.*

**History.**
**On March 1, 2009, Dubrovinskaia et al. sent a Comment to Nature, criticizing the paper by Oganov et al. (*Nature* 453, 863-867 (2009)). Their Comment was based on conflicting interests (self-promotion) and misconceptions about chemical bonding, and was rejected following this Reply.**
**In 2010, Dubrovinskaia et al. have effectively denounced the claims of their Comment in their paper**.
**To prevent misunderstanding of their paper by people with poor knowledge of chemistry, Oganov et al. published an Addendum in Nature (*Nature* 460, 292 (2009)), explaining that chemical bonding in the new phase of boron is mixed ionic-covalent (with a greater covalent component). It is the surprisingly significant degree of ionicity (unexpected in a pure element) that makes the new phase particularly interesting.**

**Abstract.**
**The comment of Dubrovinskaia et al. is scientifically flawed. The high-pressure form of boron, discovered by Oganov et al., is indeed new and its bonding has a significant ionic character, as demonstrated in Ref. 1.**

We reported[1] γ-$B_{28}$ as a new phase of pure boron that exhibits bonding with significant ionic character; the comment by Dubrovinskaia *et al.* questioning[2] these claims is based on flawed arguments.

Dubrovinskaia *et al.* claim[2] that γ-$B_{28}$ was synthesized in 1965 by a pioneer of high-pressure synthesis of materials, R.H.Wentorf (ref. 3). While our phase could be identical to Wentorf's, he only presented a qualitative X-ray diffraction pattern, which matches our phase only partially. Wentorf performed neither structure solution nor chemical analysis (which is crucial, in view of extreme sensitivity of boron structures to impurities). Wentorf's material was commonly believed not to be pure boron and was deleted years ago from the Powder Diffraction File database.

Our experiments[1] used 99.9999% pure boron and chemically inert BN capsules. Samples of γ-$B_{28}$ were shown to contain no impurities within experimental detection limits. We established[1] γ-$B_{28}$ as a pure boron phase and discovered its structure and surprising ionicity[1].

Ionicity results from charge transfer (CT) between atoms possessing different electronegativities, and significant CT was unexpected in elemental crystals. Yet, with structural formula $(B_2)^{\delta+}(B_{12})^{\delta-}$, γ-$B_{28}$ has sizable CT (δ~0.5) due to the difference of properties of the constituent $B_2$ and $B_{12}$ clusters. Ref.1 focuses on the unexpectedly high degree of ionicity in this mixed ionic-covalent element. Dubrovinskaia et al. misinterpret our atomic charges in terms of complete ionicity of γ-$B_{28}$.

Dubrovinskaia *et al.* make[2] inappropriate use of well-known concepts[4] such as CT, sublattice analysis, site-projected DOS, Bader charges, ELF. They cite high ELF values as evidence against partial ionicity, yet partially ionic GaAs and BN (with electronegativity differences of 0.4 and 1.0, respectively) have high ELF peaks on bonds (0.89 and 0.93).

Dubrovinskaia *et al.* focus[2] on the non-ionic B1-B4 bond (where both atoms bear positive charges[1]), instead of the significantly ionic B1-B2 bond. The B1-B2 distance (1.9 Å), which they describe as non-bonding, clearly corresponds to a two-electron 3-centre (B1-B2-B2) bond. Such bonds (typical distances 1.75-2.0 Å) are essential in boron chemistry[5].

Our conclusions[1] are also supported by the impact of ionicity on physical properties, e.g.: "the LO and TO modes at the zone centre are *nondegenerate* for *ionic* crystals…, whereas they are *degenerate* for *non-ionic (*homopolar) crystals" (ref. 6). The LO-TO splitting parameter ζ (Ref. 2) equals 1 for strictly covalent solids, only 1.01 for practically fully-covalent α-$B_{12}$, but rises to 1.16 for γ-$B_{28}$ (1.18 for GaAs).

Superhardness of γ-$B_{28}$, discovered subsequently[7], does not contradict partial ionicity: partially ionic cubic BN[8] and $BC_2N$[9] are also superhard. Even predominantly ionic "cotunnite-type $TiO_2$ must be among the hardest known polycrystalline materials", according to Dubrovinsky and Dubrovinskaia themselves[10].

In closing, we note that Dubrovinskaia *et al.* described in a number of their own subsequent publications γ-$B_{28}$ as a new phase[11,12] and state that "the x-ray powder diffraction pattern and the Raman spectra of the new material do not correspond to those of any known boron phases" (ref. 11). Indeed, γ-$B_{28}$ is new and significantly ionic[1].